\documentclass[a4paper,fleqn,usenatbib]{mn2e}
\usepackage[T1]{fontenc}
\usepackage{ae,aecompl}
\usepackage {graphicx,subfigure,times,aas_macros}

\newcommand{\Equ}[1]{Eq.~(\ref{eq:#1})}

\newcommand{\se}[1]{\S\ref{sec:#1}}
\newcommand{\fig}[1]{Fig.~\ref{fig:#1}}

\newcommand{\Fig}[1]{Figure~\ref{fig:#1}}

\newcommand{\tab}[1]{Table~\ref{tab:#1}}
\newcommand{\be}{\begin{equation}}
\newcommand{\ee}{\end{equation}}
\newcommand{\bea}{\begin{eqnarray}}
\newcommand{\eea}{\end{eqnarray}}

\newcommand{\msun}{{\rm M}_\odot}
\newcommand{\Msun}{M_\odot}

\newcommand{\ifm}[1]{\relax\ifmmode#1\else$\mathsurround=0pt #1$\fi}
\newcommand{\kms}{\ifmmode\,{\rm km}\,{\rm s}^{-1}\else km$\,$s$^{-1}$\fi}

\newcommand{\kpc}{\,{\rm kpc}}

\newcommand{\Gyr}{\,{\rm Gyr}}

\newcommand{\ltsima}{$\; \buildrel < \over \sim \;$}
\newcommand{\lsim}{\lower.5ex\hbox{\ltsima}}
\newcommand{\gtsima}{$\; \buildrel > \over \sim \;$}
\newcommand{\gsim}{\lower.5ex\hbox{\gtsima}}

\def\la{\langle}

\def\sy{\,M_\odot\, {\rm yr}^{-1}}

\def\cmc{\,{\rm cm}^{-3}}
\def\cms{\,{\rm cm}^{-2}}

\def\Ms{M_*}

\newcommand{\Halpha}{H${\alpha}$}
\def\re{r_e}

\usepackage{color}

\large 

\title[Outflows]
{Velocities of Warm Galactic Outflows from Synthetic \Halpha \ Observations of Star-forming Galaxies} 

\author[Ceverino et al.]
{\parbox[t]{\textwidth} 
{ 
Daniel Ceverino$^{1,2,5}$\thanks{E-mail: daniel.ceverino@cab.inta-csic.es},
Santiago Arribas$^{1,2}$,
Luis Colina$^{1,2}$, 
Bruno Rodr{\'i}guez Del Pino$^{1,2}$, \\
Avishai Dekel$^3$,
Joel Primack$^4$}
\\ \\
$^1$Centro de Astrobiolog{\'i}a (CSIC-INTA), Ctra de Torrej{\'o}n a Ajalvir, km 4, E-28850 Torrej{\'o}n de Ardoz, Madrid, Spain \\
$^2$Astro-UAM, Universidad Autonoma de Madrid, Unidad Asociada CSIC, E-28049 Madrid, Spain \\
$^3$Center for Astrophysics and Planetary Science, Racah Institute of Physics, The Hebrew University, Jerusalem 91904, Israel \\
$^4$Department of Physics, University of California, Santa Cruz, CA,
95064, USA \\
$^5$ Zentrum fur Astronomie der Universitat Heidelberg, Institut fur
Theoretische Astrophysik, Albert-Ueberle-Str. 2, 69120 Heidelberg, Germany
}

\date{}

\pubyear{2015}

\begin{document}
\label{firstpage}
\pagerange{\pageref{firstpage}--\pageref{lastpage}}
\maketitle

\begin{abstract}
The velocity structure imprinted in the \Halpha \ emission line profiles contains valuable information about galactic outflows.
Using a set of high-resolution zoom-in cosmological simulations of galaxies at $z\simeq2$, 
we generate \Halpha \ emission line profiles, taking into account the temperature-dependent \Halpha \ emissivity, as well as dust extinction. 
The \Halpha \ line can be described as a sum of two gaussians, as typically done with observations.
In general, its properties are in good agreement with those observed
in local isolated galaxies with similar masses and star formation
rates, assuming a spatially constant clumping factor of $c\simeq24$.
Blueshifted outflows are  very common in the sample.
They extend several kpc above the galaxy discs.
They are also spread over the full extent of the discs.
However, at small radii, the material with high velocities tends to remain confined within a thick disc, as part of galactic fountains or a turbulent medium, most probably due to the deeper gravitational potential at the galaxy center.
\end{abstract}

\begin{keywords} 
galaxies: evolution --- 
galaxies: formation  
\end{keywords} 

\section{Introduction}
\label{sec:intro}

The study of star-forming (SF) galaxies provide relevant insights about the key mechanisms in the formation of galaxies.
In particular, many SF galaxies show broad emission or absorption lines in their spectra, indicating high velocities relative to the systemic or mean velocity of the galactic interstellar medium (ISM). 
These high velocities are usually interpreted as strong galaxy outflows.
However, many of the properties of these outflows, like their extent beyond the ISM or their spread within the galaxy, remain poorly understood.

Massive star formation and/or an active galactic nucleus (AGN) would expel the surrounding interstellar medium as a consequence of the mechanical and radiative energy liberated, generating strong outflows \citep{DiMatteo05, OstrikerShetty11, Murray11, Hopkins12}. 
The role of these outflows governing the subsequent galaxy evolution is believed to be crucial, as they can regulate and quench both star formation and black hole activity through feedback loops.
Outflows are also the primary mechanism by which dust and metals are redistributed over large scales within the galaxy, or the circumgalactic medium (CGM) \citep{Steidel10}, or even expelled into the intergalactic medium (IGM)  \citep[][for a review]{Veilleux05}.  
Hence most of the models of galaxy formation require energetic outflows to reproduce the observed properties of massive galaxies \citep{DekelSilk86, SilkRees98, DiMatteo05, Hopkins06, Hopkins12}.  
Although this general theoretical framework is reasonably well established, the relative role of the different feedback mechanisms, driven by stars or AGNs,  is not fully understood. 
Also, it is not well determined how the outflow velocities depend on the mass of the host galaxy, its star formation rate, and/or the type and luminosity of the AGN.

Therefore the study of outflows at high-z has gained much interest in recent years.  A remarkable observational result is that outflows are very prominent and ubiquitous in SF galaxies during  the peak of the Universe star formation history at redshifts $z\sim$ 1-3 \citep{Shapley03, Weiner09, Steidel10, Alexander10, Maiolino12, Cano-Diaz12}. 
Recent studies are starting to characterize ionized gas outflows at high-z with samples covering different  stellar mass ranges. \citep[e.g.][]{Shapiro08, Harrison12, Swinbank12}. For the most massive galaxies (i.e. $> 10^{11} \Msun$) outflows seem to be launched by powerful nuclear AGNs \citep{ForsterSchreiber14, Genzel14}, while in less massive systems the starburst is likely to dominate their output energy budget \citep{Newman12}.  

Locally, (Ultra)Luminous Infra-Red Galaxies (LIRGs and ULIRGs)  exhibit the most conspicuous cases for gas outflows in the local universe. They form stars at rates similar to those of main-sequence star-forming galaxies at $z\sim2-3$, offering the possibility to study the outflow phenomenon at much higher S/N and linear resolutions than at high-z. Consequently, a number of studies have been aimed at  characterizing their properties \citep[][and references therein]{Veilleux05}.   

The ionized outflows at low and high redshifts are generally studied using as observational tracers the strong optical emission lines, like \Halpha \ \citep[e.g.][]{Shapiro09, Newman12, ForsterSchreiber14, Genzel14, Westmoquette12, RodriguezZaurin13, RupkeVeilleux13, Bellocchi13, Arribas14, Rich15}.

Theoretical predictions using synthetic observations in \Halpha \ provide a valuable and fair comparison with existing and future observations in \Halpha.
Cosmological simulations of galaxy formation  have shown the kinematics of ionized gas in violently unstable but rotating discs at $z\sim2$ \citep{Ceverino12, Genel12}.
These discs show strong and turbulent, non-ordered motions of the order of 40-50 $\kms$, in addition to ordered rotation of the order of 200 $\kms$.
However, these studies have focused on the kinematics of the dense, \Halpha-bright ISM, by measuring the mean velocity and dispersion at different galaxy positions assuming a single profile component.
 Little attention has been paid to the kinematics of the diffuse, \Halpha-faint gas (the tail of the \Halpha \ line), whose kinematics properties could be radically different.

In this paper we present synthetic \Halpha \ observations based on Adaptive Mesh Refinement (AMR) simulations of galaxy formation. In this initial study we focus on the stellar mass range of $\Ms \simeq$$10^{10} \Msun$, and ignore the effects of the AGN.  
Therefore, we focus on the properties of stellar-driven outflows in moderately massive SF galaxies with ${\rm SFR}\sim30 \sy$,
i.e., typical of local LIRGs \citep{Bellocchi13, Arribas14}. 
%
%
We will compare the predicted kinematic properties of the warm, \Halpha \ emitting gas with  observations, looking at characteristics generally associated with outflows, like high velocities. To this aim we will follow with our synthetic observations a similar methodology  as typically used with observed data. In particular we will fit a 2-Gaussian model to the synthetic \Halpha \ line profile, in order to identify a broad component, which is generally associated to the outflow. Then, we will compare with that derived from the data, and discuss the validity of that approach.  

In this first approach, we are focused on some basic properties of the outflows, such as their spatial extend outside the galaxy or their spread within the disc. 
Another important property, relevant for the interpretation of observations, is the effect of the dust attenuation on the outflows velocities.
The paper is structured as follows.
In section \se{runs}, we describe the simulation data.
\se{Halpha} describes the modeling of the \Halpha \ emission.
\se{maps} shows the synthetic images in \Halpha.
\se{Vz} uses the distribution function of vertical velocities as a proxy for the \Halpha \ emission line profile around the galactic systemic velocity.
\se{profiles} focuses on the vertical extent of the outflow perpendicular to the galaxy plane.
\se{nuclear} compares the circumnuclear outflows with the galaxy-wide  outflows.
\se{dust} discusses the effect of  increasing dust attenuation on the outflows properties.
Finally, \se{conclusion} is devoted to discussion and summary.

\begin{table} 
\caption{Properties of the zoom-in simulations. Columns show name of the run, redshift, virial radius, virial mass, stellar mass within 0.1$R_{\rm vir}$ and SFR in $\sy$. Masses are in $\Msun$ and radii in kpc.}
 \begin{center} 
 \begin{tabular}{cccccc} \hline 
\multicolumn{2}{c} {Run } \ \ \ z \ \ \ & $R_{\rm vir}$ & virial mass & stellar mass  & SFR \\
\hline 
V06   & 2.0 & 88 & $5.4 \times 10^{11}$  & $2.1  \times 10^{10}$ & 21 \\
V07   & 2.7 & 80 & $7.2 \times 10^{11}$  & $3.7  \times 10^{10}$ & 61 \\
V08   & 1.2 & 80 & $6.8 \times 10^{11}$  & $2.7  \times 10^{10}$ & 28 \\
V12   & 3.5 & 44 & $2.3 \times 10^{11}$  & $1.3  \times 10^{10}$ & 16 \\
V13   & 1.6 & 99 & $5.0 \times 10^{11}$  & $1.9  \times 10^{10}$ & 30 \\
V19   & 4.0 & 47 & $3.6 \times 10^{11}$  & $2.3  \times 10^{10}$ & 32 \\
V22   & 3.2 & 60 & $4.4 \times 10^{11}$  & $2.5  \times 10^{10}$ & 31 \\
V26   & 2.7 & 80 & $3.2 \times 10^{11}$  & $1.3  \times 10^{10}$ & 17 \\
V27   & 1.8 & 82 & $3.7 \times 10^{11}$  & $1.1  \times 10^{10}$ & 21 \\
V32   & 3.0 & 55 & $3.0 \times 10^{11}$  & $1.6  \times 10^{10}$ & 21 \\
V33   & 2.7 & 66 & $4.2 \times 10^{11}$  & $1.5  \times 10^{10}$ & 32 \\
V34   & 2.1 & 81 & $4.7 \times 10^{11}$  & $1.3  \times 10^{10}$ & 21 \\
 \end{tabular} 
 \end{center} 
\label{tab:1} 
 \end{table} 

 \begin{figure}
\includegraphics[width=0.49 \textwidth]{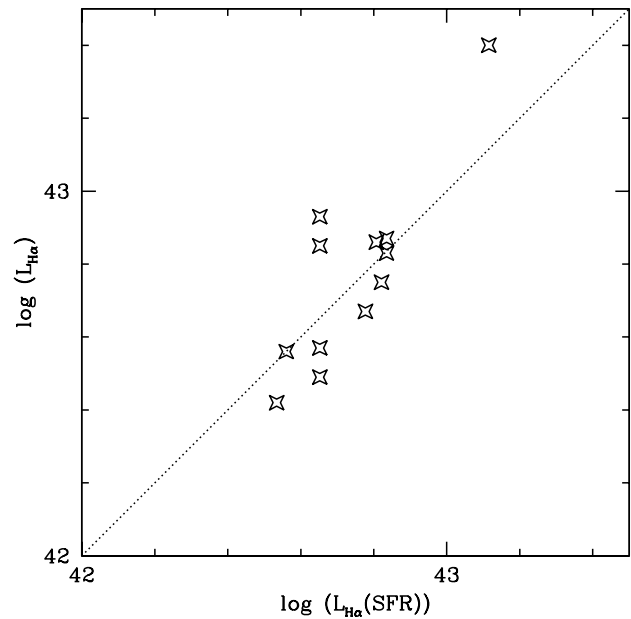}
\caption{Comparison between the \Halpha \ luminosity of each galaxy according to its SFR \citep{Kennicutt98} and the \Halpha \ luminosity computed according to section \se{Halpha}. }
\label{fig:Lalpha}
\end{figure}

\begin{figure*}
\includegraphics[width=0.99 \textwidth]{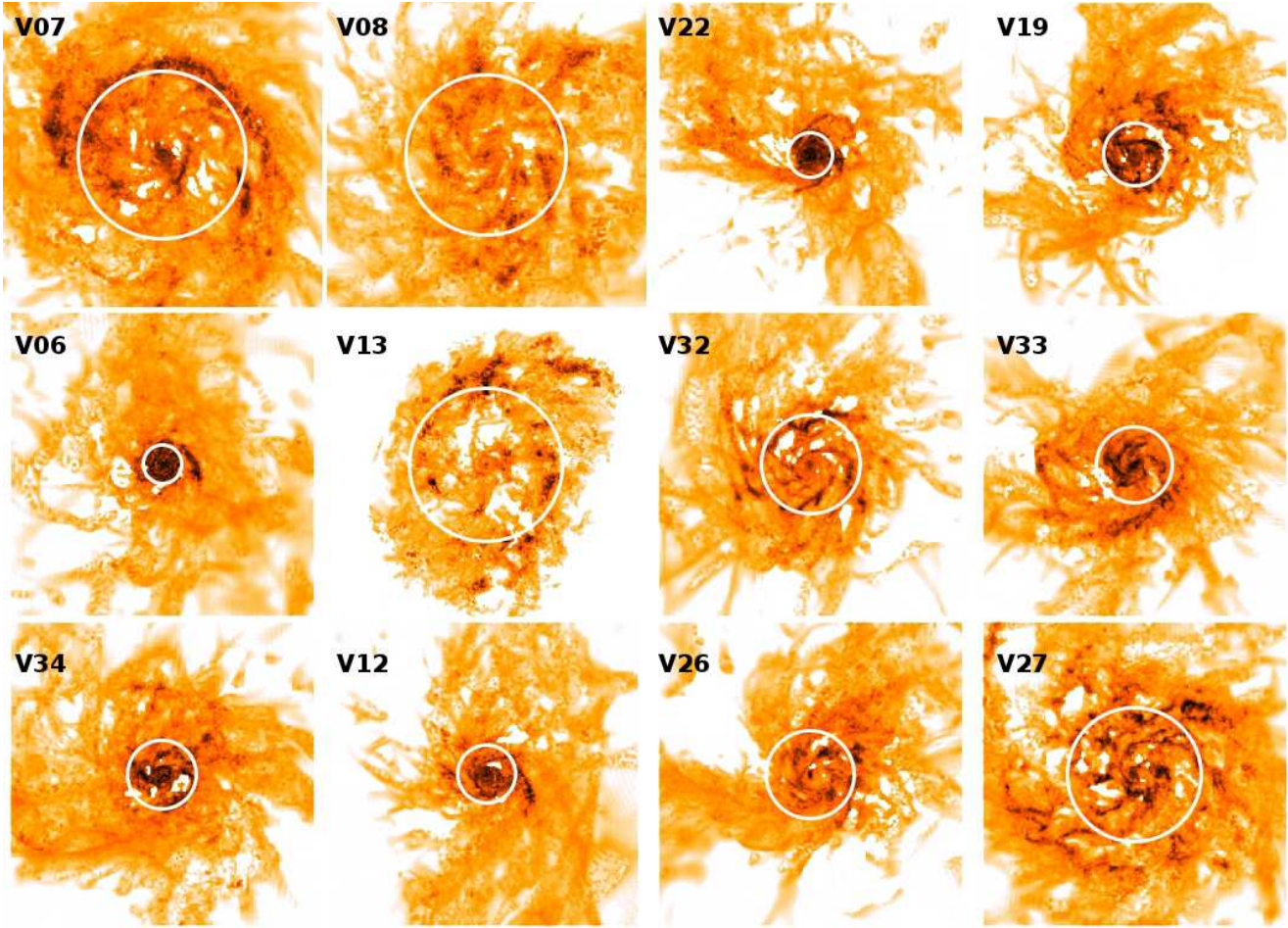}
\caption{Synthetic \Halpha \ images ordered by decreasing stellar mass. The size of the images is $20 \times 20 \kpc^2$.  White circles mark the \Halpha \ effective radius ($\re$).
\Halpha-bright clumps are distributed in the galaxy discs along rings and floculent spiral arms. 
The discs are also embedded in a diffuse and warm medium that also emits \Halpha.}
\label{fig:maps}
\end{figure*}

 \begin{table*} 
\caption{Properties of the \Halpha \ synthetic observations. Columns show name of the run, \Halpha \ effective radius in kpc ($\re$), flux ratio between the broad and narrow components of the velocity distribution (F(B)/F(N)), and maximum velocity of the outflow ($V_{\rm max}$) in km s$^{-1}$ for the whole galaxy (2$\re$) as defined in Arribas et al. (2014),  F(B)/F(N) and  $V_{\rm max}$ for the circumnuclear outflows (0.5$\re$), and F(B)/F(N) and $V_{\rm max}$ for the galaxy outflows assuming five times more dust attenuation.}
 \begin{center} 
 \begin{tabular}{cccccccc} \hline 
\multicolumn{2}{c} {Run } \ \ \ $\re$ & F(B)/F(N) & $V_{\rm max}$ & (F(B)/F(N))$_{\rm nuclear}$ & $(V_{\rm max})_{\rm nuclear}$ & (F(B)/F(N))$_{\rm more dust}$ & $(V_{\rm max})_{\rm more dust}$ \\
\hline 
V06   & 1.2 & 0.25 & 120 & 0.71 & 40 & 0.65 & 120\\
V07   & 6    & 0.54 & 140 & 2.7 & 160 & 0.5 & 120 \\
V08   & 5.2 & 0.25 & 99  & 0.16 & 80 & 0.43 & 105 \\
V12   & 1.9 & 0.19 & 120 & 0.57 & 120 & 0.27 & 130\\
V13   & 7.5 & 0.43 & 80 & 0.5 & 50 & 1.4 & 80 \\
V19   &  2   & 0.30 & 133 & 0.26 & 130 & 0.45 & 130\\
V22   & 1.4 & 1.3 & 128 & 0.83 & 110 & 2.2 & 120 \\
V26   & 2.9 & 0.19 & 140 & 0.24 & 140 & 0.67 & 106\\
V27   & 5    & 0.18 & 100 & 0.17 & 100 & 0.26 & 110 \\
V32   & 3.2 & 0.3 & 70 & 0.36 & 40 & 0.81 & 80 \\
V33   & 2.5 & 4.9 & 101 & 4.9 & 90 & 11 & 80 \\
V34   & 2.2 & 0.32 & 80 & 0.33 & 80 & 0.53 & 84\\
 \end{tabular} 
 \end{center} 
\label{tab:2} 
 \end{table*} 
 
 \begin{figure*}
\includegraphics[width=0.99 \textwidth]{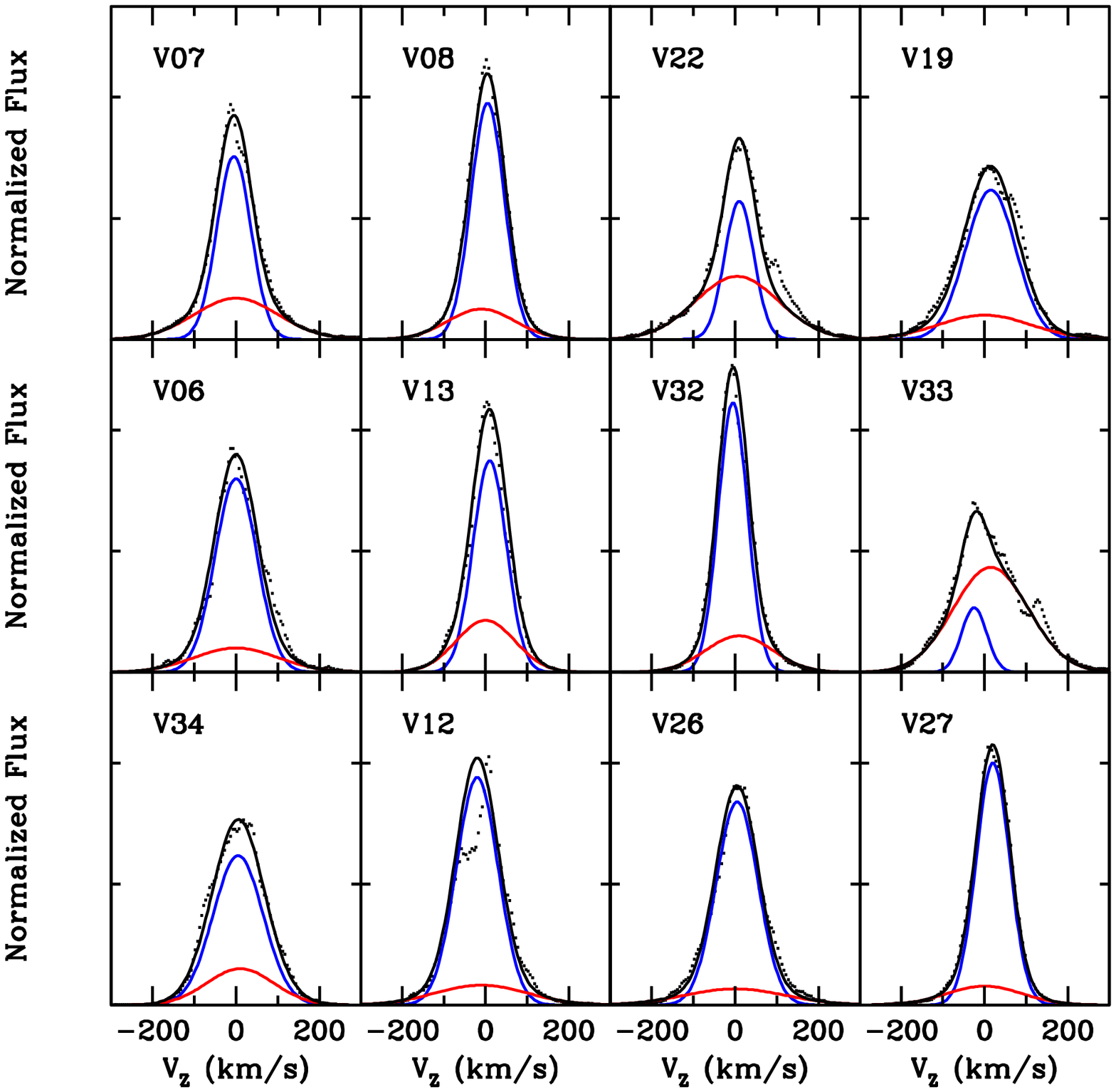}
\caption{Distribution functions of the vertical velocities ($V_z$)
  perpendicular to the galaxy plane, weighted by \Halpha \ flux (black
  points). All the \Halpha-emitting gas inside 2$\re$ is considered. 
This is a proxy for the \Halpha \ emission line profile, integrated over the whole galaxy.
A narrow gaussian plus a broad gaussian fit the total distribution. 
The broad component could be identity as an outflow. }
\label{fig:Vz_dust}
\end{figure*}

\section{Simulations}
\label{sec:runs}

 %
 The zoom-in simulations used in this paper were drawn from a larger dataset, the  \textsc{Vela} sample 
 \citep{Ceverino14, Zolotov}.
 We select for each simulation the latest snapshot that fulfills the following criteria:
\begin{itemize}
\item
A galaxy stellar mass of $\Ms \ge 10^{10} \ \Msun$,
\item
A specific star-formation-rate of $sSFR \ge 1 \Gyr^{-1}$,
\end{itemize}
where $sSFR=SFR/\Ms$ (\tab{1}).
Therefore, we select SF galaxies with a narrow mass range: $\Ms=(1-4) \times 10^{10} \ \Msun$ and $SFR=(20-60) \sy$, and a mean redshift of $z=2.5 \pm 0.8$.
These are the typical values of moderately massive main-sequence galaxies at $z=2$ \citep{Whitaker14, Wisnioski15}, and  local LIRGs \citep{Bellocchi13, Arribas14}.
Ongoing major mergers or young remnants (150-200 Myr after coalescence) usually have 
strong gas motions, related to merger interactions and tidal forces \citep{Colina05}, so they are excluded from the analysis.

%
The simulations were performed with the  \textsc{ART} code
\citep{Kravtsov97,Kravtsov03}, which accurately follows the evolution of a
gravitating N-body system and the Eulerian gas dynamics using an AMR approach.
Beyond gravity and hydrodynamics, the code incorporates 
many of the physical processes relevant for galaxy formation.  
These processes, representing subgrid 
physics, include gas cooling by atomic hydrogen and helium, metal and molecular 
hydrogen cooling, photoionization heating by a constant cosmological UV background with partial 
self-shielding, star formation and feedback, as described in 
\citet{Ceverino09}, \citet{CDB}, and \citet{Ceverino14}. 
 In addition to thermal-energy feedback, the simulations use radiative feedback.
This model adds a non-thermal pressure, radiation pressure, to the total gas pressure in regions where ionizing photons from massive stars are produced and trapped. 
In the current implementation, named RadPre in \citet{Ceverino14}, radiation pressure is included in the cells (and their closest neighbors) that contain stellar particles younger than 5 Myr and whose gas column density exceeds $10^{21}\ \cms.$ 

 The initial conditions of these runs contain from 6.4 to 46 $10^6$ dark matter particles 
 with a minimum mass of 
$8.3 \times 10^4 \ \msun$, while the particles representing single stellar populations that were formed in the simulation
have a minimum mass of $10^3 \ \msun$. 
The maximum spatial resolution is between 17-35 proper pc. More details can be found in \citet{Ceverino14} and \citet{Zolotov}.

\section{Modeling emission in \Halpha}
\label{sec:Halpha}

The first step is to compute, per each cell of the simulation, the emission line coefficient for \Halpha,
\begin{equation} 
j_{ {\rm H}\alpha } = (4 \pi)^{-1}  f_{\rm HII}^2 \ c^2 \ n_{\rm H}^2 \ \alpha^{\rm eff}_{{\rm H}\alpha} \  h  \nu^{\ \ }_{{\rm H}\alpha} \  [ {\rm erg} \ {\rm s}^{-1} {\rm cm}^{-3}]
\label{eq:Ha}
\end{equation} 
where $f_{\rm HII}$ is the hydrogen ionization fraction, taken from the same  \textsc{cloudy} tables used for cooling \citep{Ceverino09},
 $n_{\rm H}$ is the hydrogen number density, $h \nu^{\ \ }_{{\rm H}\alpha}$ is the energy of a single \Halpha \ photon and $\alpha^{\rm eff}_{{\rm H}\alpha} $ is the temperature-dependent effective recombination coefficient, tabulated assuming case B recombination in the low density limit \citep{Osterbrock}.
This assumption is valid for the density regime of these simulations ($n_{\rm H} < 10^4 \cmc$), where collisional effects on the strength of the recombination lines are low, within 10\%.

\subsection{\Halpha \ Luminosities and Clumping Factor}

The origin of the \Halpha-emitting gas is still unclear, specially within the outflowing material.
Previous works argued that the high-velocity, \Halpha \ emission could arise in clouds or filaments that formed as a result of fragmentation of clouds embedded in the high-velocity outflow \citep{Fujita09} or as a result of swept-up remnants of cold clouds that existed in the ISM \citep{Cooper08, Cooper09}.
These works indicate that \Halpha \ is emitted from structures that are much smaller than the spatial resolution of the simulations presented in this paper. Therefore, the densities in these filaments could be significantly higher than the density averaged over a single resolution element. This will affect the \Halpha \ emissivities (\Equ{Ha}).
As often done in the literature  \citep{Gnedin97} we introduce a clumping factor, $c$, that accounts for this difference in densities.

However, the value and variability of this clumping factor is unclear, due to the complex interplay between processes at different scales, such as gravity, hydrodynamics, and radiation. We decide to assume a constant value for the clumping factor because we are interested in the shapes of the \Halpha \ line, not in its absolute value and we constrain the clumping factor by using observations. 

In order to test this assumption, 
\Fig{Lalpha} shows,in the x-axis, the \Halpha \ luminosity using the empirical dependence with the SFR, measured from the simulations (\tab{1}), according to \cite{Kennicutt98},
\begin{equation} 
 \log {\rm L}_{\rm H\alpha} = \log {\rm SFR} + 41.33 . 
 \end{equation} 
The y-axis shows the \Halpha \ luminosity assuming a clumping factor of $\sim$24 (this means that the typical densities that emit \Halpha \ photons are 24 times higher than the densities computed in the simulations, coming along with the densities measured in \cite{Arribas14}).  This value was computed as follows. Per each run, we compute the clumping factor needed to match the above relation. Then, we take the average, $c=24.3 \pm$ 3.8. Its 1-$\sigma$-scatter is only 15\%, which seems small given the relatively large range of galaxy and outflows morphologies (SFR ranges between 16-60 $\sy$). Therefore, our assumption of a constant clumping factor seems reasonable. Its actual value does not modify the results of this paper, which are based on normalized fluxes, not absolute luminosities.

\subsection{Dust attenuation}

We also consider the attenuation of the emission in each cell due to the intervening dust along the line-of-sight.
First, the dust column density was computed using the metal-to-dust conversion used in \cite{Snyder15}.
Then we take the dust opacity at $\lambda$=6563 \AA \ from the model of MW diffuse HI clouds of \cite{WeingartnerDraine01}, including a factor of 2 larger opacity in the \Halpha \ line with respect to the continuum \citep{Calzetti01}.
This results in attenuations of $A_v\simeq1-2$, typical of local isolated LIRGs \citep{Veilleux95}.
Due to the uncertainties in all these conversions at high-z, we have redone the analysis using 5 times more attenuation (See \se{dust}). 

\section{Synthetic images in \Halpha}
\label{sec:maps}

\begin{figure}
\includegraphics[width=0.49 \textwidth]{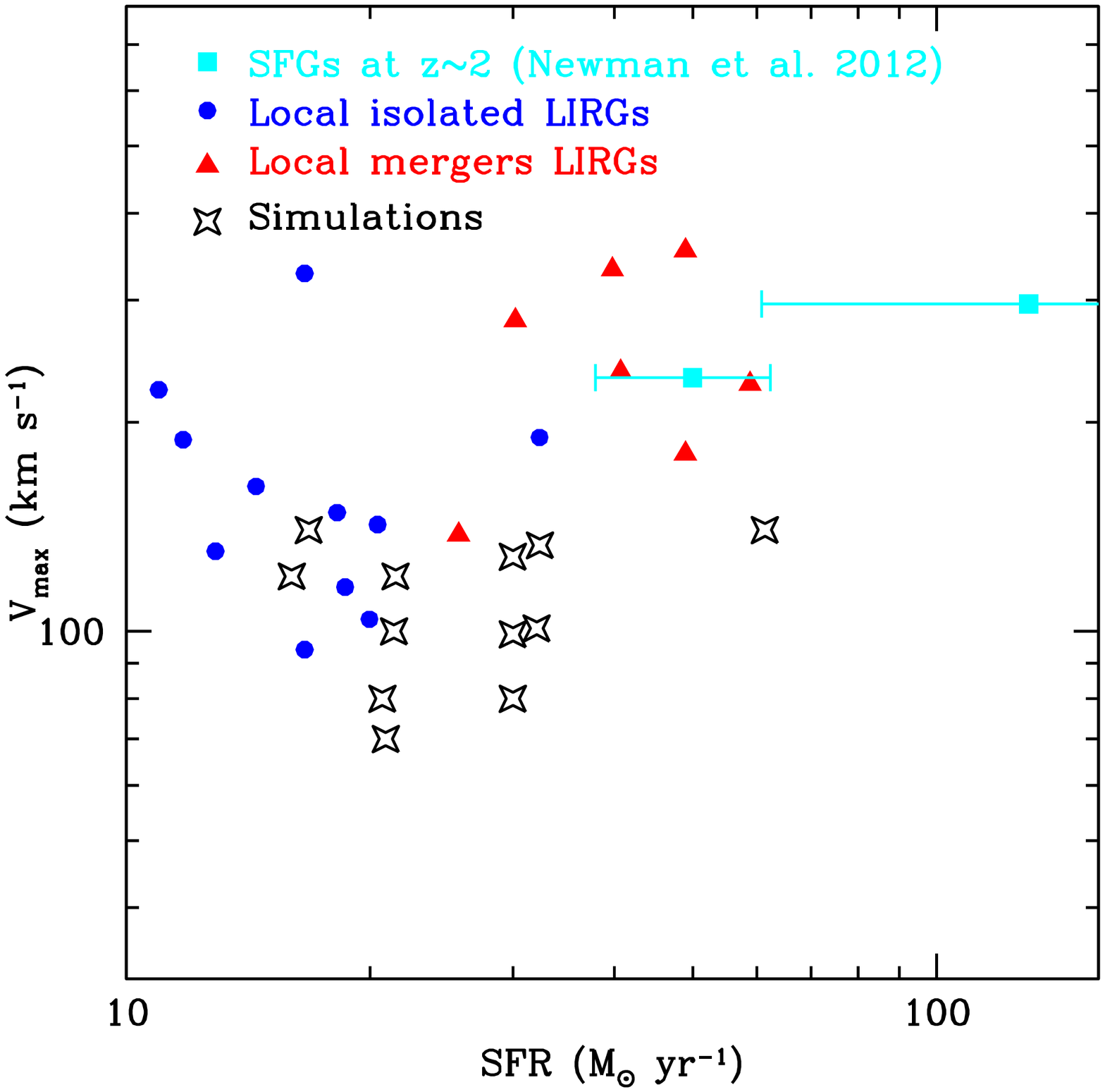}
\includegraphics[width=0.49 \textwidth]{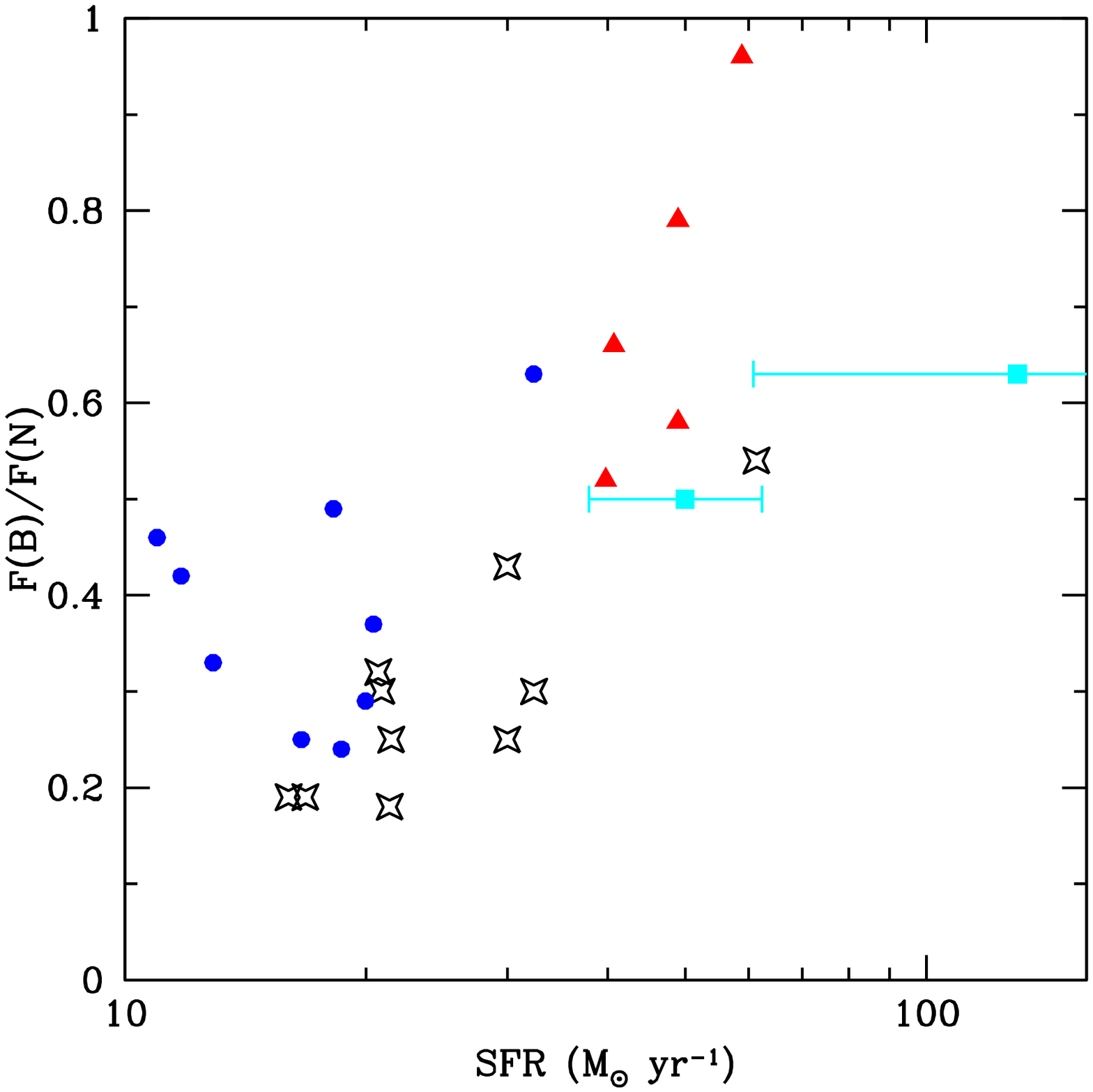}
\caption{Top: Maximum velocity ($V_{\rm max}$) versus SFR for the simulated galaxies, the local sample of isolated and merger LIRGs of Arribas et al. (2014) and the stacks of $z\simeq2$ SF galaxies from Newman et al. (2012). Bottom: Broad-to-narrow flux line ratio versus SFR. The simulations are consistent with the subsample of isolated LIRGs.}
\label{fig:Vmax}
\end{figure}

\begin{figure*}
\includegraphics[width=0.99 \textwidth]{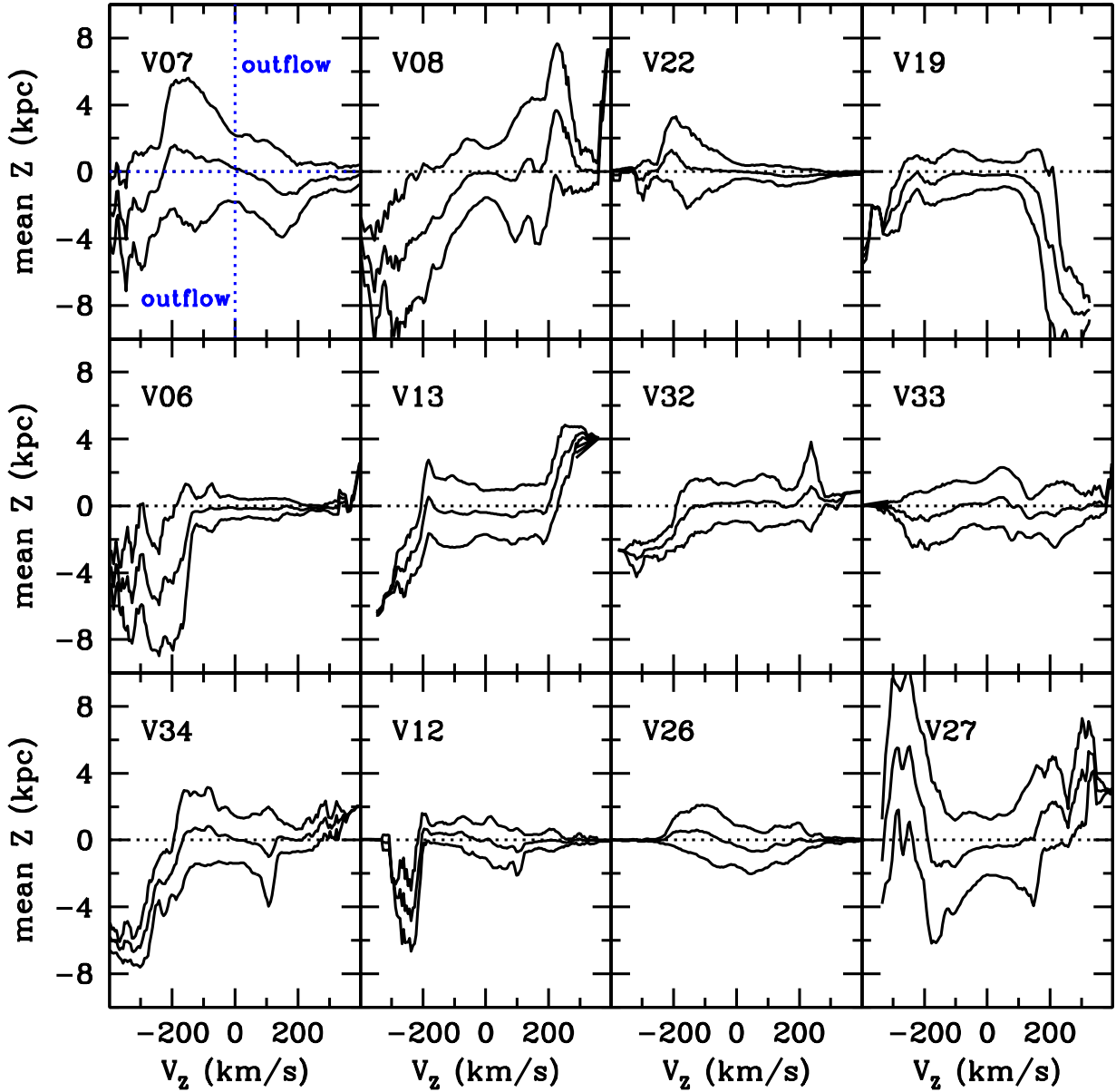}
\caption{\Halpha-weighted mean height and $\pm\sigma$ dispersion in the near-side of the galaxy (Z<0) or the far-side (Z>0) for the velocity bins used in \Fig{Vz_dust}.
Large-scale outflows can be identify with highly negative (blueshifted) velocities coming from the near-side of the galaxy.
They extend upto 8 kpc away from the disc plane.}
\label{fig:Z_dust}
\end{figure*}

The next step is to integrate the \Halpha \ emissivity, attenuated by dust along a given line-of-sight.
\Fig{maps} shows the synthetic \Halpha \ images in the face-on view, defined by the angular momentum of the cold ($T<10^4$K) gas within some fixed radius \citep{CDB}.
The \Halpha \ images show clumpy discs of different sizes, ranging from an effective radius $\re=6 \kpc$ (V07) to $\re=1.2 \kpc$ (V06) (\tab{2}).
Therefore, our sample includes extended and compact \Halpha \ sources.

The synthetic \Halpha \ images have morphologies and sizes similar to 
those observed in local LIRGs \citep{RodriguezZaurin11, Arribas12}, and SF main-sequence galaxies at  $z\simeq2$
\citep{ForsterSchreiber09, Genzel11, Wisnioski15}.
 \Halpha-bright clumps are distributed along rings or floculent spiral arms within the discs. 
In the irregular and turbulent ISM, there are also holes due to its multiphase nature. These holes are filled with hot and diffuse gas that does not emit significantly in \Halpha.
The mean temperature of the \Halpha \ emitting gas is around $10^4$ K, the typical temperatures of star-forming and/or ionized gas.
The discs are also embedded in a diffuse medium, composed of inflowing and outflowing material that may also emit \Halpha \ photons \citep{Ceverino15c}.

In the next sections, we will focus on the vertical velocities of this \Halpha \ emitting gas and their extent perpendicular to the galaxy plane. 
In all cases, there is strong \Halpha \ emission in the galaxy center.
However, examples like V07 and V13 show stronger emission concentrated in clumpy SF rings around $\re$.
We will see if the outflows are concentrated at the center or correlated with the distributions of clumps.

\section{Velocity distribution functions}
\label{sec:Vz}

 \begin{figure}
\includegraphics[width=0.49 \textwidth]{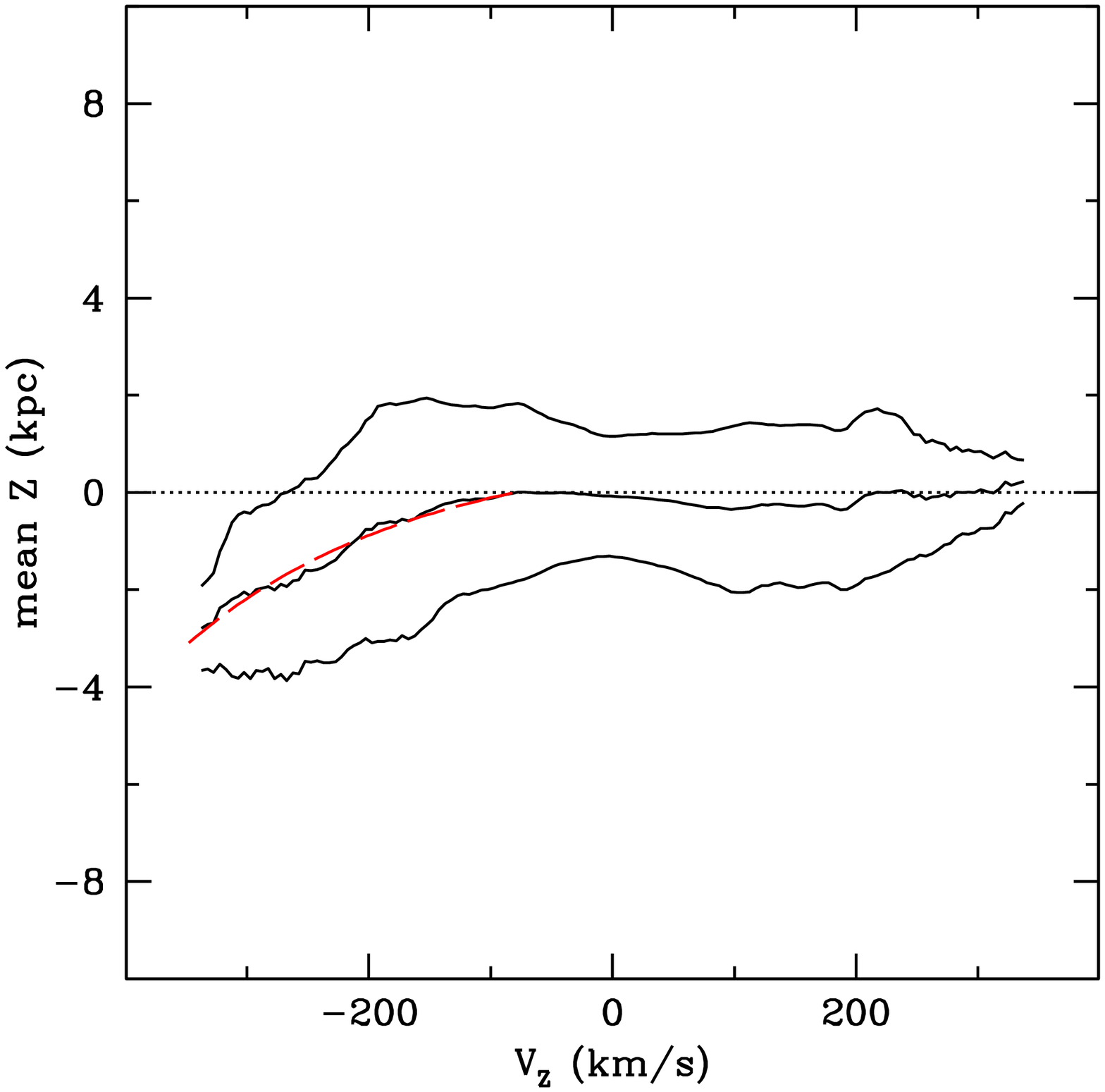}
\caption{Mean height (and dispersion) after stacking all profiles of \Fig{Z_dust}. 
The mean profile shows a blueshifted outflow plus galactic fountains with velocities of $\pm200 \kms$. 
The mean height of the outflow can be fitted by an exponential profile (red dashed curve).}
\label{fig:Z_mean}
\end{figure}

\begin{figure}
\includegraphics[width=0.49 \textwidth]{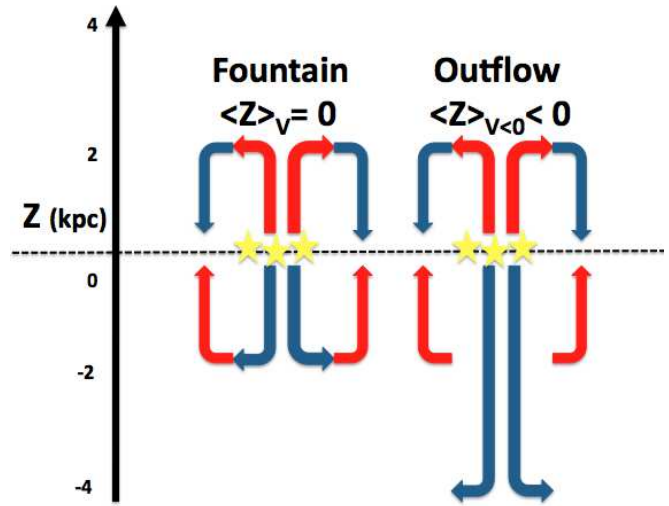}
\caption{Cartoon of a galactic fountain plus a galactic outflow. Red arrows represent material moving with positive vertical velocities (redshifted) while blue arrows  have negative velocities (blueshifted material). The observer is at $Z=-\infty$. In the fountain scenario, material with the same high velocity is moving at both sides of the galaxy plane ($Z=0$), therefore its mean height is zero. On the other hand, a blueshifted outflow biases the mean height towards the near-side of the galaxy ($Z<0$). }
\label{fig:fountain}
\end{figure}

\begin{figure*}
\includegraphics[width=0.99 \textwidth]{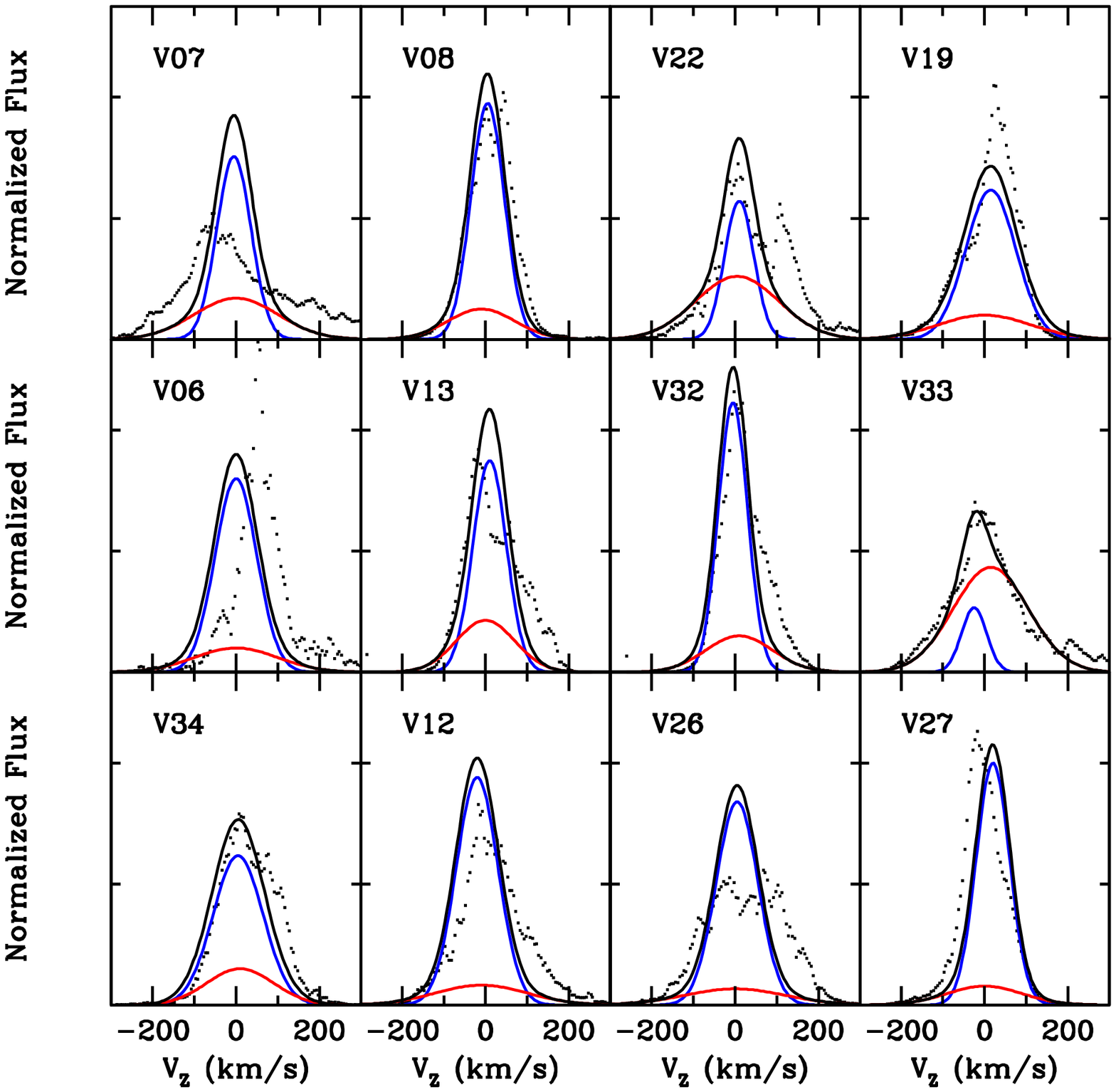}
\caption{Velocity distribution functions inside $0.5\re$ (Black points). Curves are the same as in \Fig{Vz_dust}.
The distributions are broader, so the contribution of the broad component is higher than in the case of the whole galaxy.
However, the maximum velocities of the outflows are similar.
It seems that the outflows velocities are similar throughout the galaxy.}
\label{fig:Vz_nuclear_dust}
\end{figure*}

We want to disentangle inflows and outflows from the internal gas motions within the interstellar medium of these simulated galaxies.
Therefore, we select an aperture including the whole galaxy (2 $\re$) in the face-on view (\Fig{maps}) and compute the vertical velocity distribution function within that radius, weighted by the \Halpha \ flux.
This is a proxy for the \Halpha \ emission line profile around the galactic systemic velocity.
By selecting a face-on view, rotational velocities as well as any other gas motions along the galactic plane are ignored.
In these numerical experiments, we will concentrate in the characterization of the velocities of warm, \Halpha-selected outflows perpendicular to the galactic disc.
The velocity distributions (\fig{Vz_dust}) are well fitted by the sum of two gaussians: narrow and broad components.
This approach  is commonly followed to interpret the observed velocity distributions \citep[e.g.][and references therein]{Genzel11, Arribas14}.

The central velocity of the narrow component is consistent with the systemic velocity within uncertainties ($5 \pm 10 \kms$). 
The width of the narrow component, $\sigma_{\rm N}=40 \pm 10 \kms$, is higher than
those of local main-sequence star-forming galaxies, $\sigma < 25 \kms$ \citep{Epinat10},  for which typical SFRs are  $\sim 1 \sy$.
However, the simulated width is in good agreement with the sub-sample of 11 isolated local LIRGs without AGN traces,  observed by \cite{Arribas14} with a median value of $\sigma=42 \kms$.
It is also consistent with the values found in $z\simeq2$ galaxies of similar masses \citep{Wisnioski15}.
These authors identify this component with the internal (turbulent) motions within the violently unstable discs.

The broad component has a significantly higher dispersion than the narrow component, $\sigma_{\rm B}=95 \pm 18 \kms$.
In almost all cases, velocities higher than 200 km s$^{-1}$ are observed at either sides of the systemic velocity.
This could indicate the presence of outflows in all the galaxies of the sample.
The contribution of the broad component to the total line profile is smaller than the narrow component. 
Excluding two extreme cases (V22 and V33), the median of the broad-to-narrow flux ratio is F(B)/F(N)=$0.3 \pm 0.1$  (\tab{2}).
This means that the broad component is roughly three times smaller than the narrow component, indicating that it represents a more diffuse material with lower \Halpha \ emissivity.

There are little differences between the velocity distributions of the different galaxies.
Both compact and extended \Halpha \ sources show similar distributions.
Two of the least massive galaxies ($\Ms \simeq 10^{10} \ \msun$) show the smallest broad components, (F(B)/F(N) $\leq 0.2$ for V26 and V27).
On the other hand, the two extreme cases where the broad component dominates (F(B)/F(N)>1) show indications of large-scale gas flows produced by a relatively recent (200-250 Myr) major merger.
Observationally, there are clear indications that the relative strength of the broad component is more important in recent or ongoing mergers \citep{Arribas14}.

As often done to interpret observations \citep[e.g.][]{Veilleux05, Rupke05} we define a "maximum" velocity of the outflow as
\begin{equation} 
V_{\rm max} = {\rm abs} ( v_{\rm offset} -{\rm FWHM}/2)  = {\rm abs} ( v_{\rm offset} - 1.18  \sigma_{\rm B} )
\end{equation} 
where $v_{\rm offset}$ and $\sigma_{\rm B}$ are the offset and dispersion of the broad gaussian component.
In the top panel of \Fig{Vmax}, these velocities are compared with the observed values from the sample of local LIRGs without AGN of \cite{Arribas14}.
The median of the simulated sample, $ V_{\rm max} = 110 \pm 20 \kms$, is slightly lower but within the 1$\sigma$ uncertainty of the observed value of $ V_{\rm max} = 150 \pm 70 \kms$ for the subsample of isolated LIRGs.
On the other hand, the subsample of LIRG mergers have significantly higher values, $ V_{\rm max} = 310 \pm 140 \kms$.
Therefore, the simulated outflows are not consistent with the outflows produced in mergers.
This result was expected because we excluded mergers from our sample of simulations. 
Future studies will target the merger scenario.


The bottom panel of \Fig{Vmax} also compares the simulated and observed values of the broad-to-narrow flux ratio (F(B)/F(N)).
The values of isolated LIRGs,  F(B)/F(N)=$0.37 \pm 0.13$ are consistent with the simulated values, described above.
These simulated values are also consistent with the fraction of \Halpha \ flux from the high-$\sigma$ components in the subsample of nearby isolated LIRGs from the WiFeS-GOALS survey \citep[][Fig. 9]{Rich15}.
We conclude that the kinematics of ionized gas (both narrow and broad components) from simulated, main-sequence galaxies at $z\simeq2$ are consistent with the kinematics of local isolated LIRGs.

Finally, we compare the simulations with observations of SF galaxies at $z\simeq2$ from stacking spectra \citep{Newman12}.
In \Fig{Vmax}, we show two stacks. 
The datapoint with high SFR corresponds to SF galaxies with stellar masses higher than $10^{10} \ \msun$. The median of the sample is $\Ms=2.2 \times 10^{10} \ \msun$, similar to the simulated sample. 
However, the median SFR is significantly higher, ${\rm SFR}\simeq130 \sy$. 
This is a factor 2.5 higher than the ridge of the main-sequence of SF galaxies at $z\simeq2$ \citep{Whitaker14, Wisnioski15}.
The second stack is composed of galaxies with ${\rm SFR} < 100 \sy$,
but the median mass of this sample is significantly lower ($\Ms=0.9 \times 10^{10} \ \msun$) 
than the median mass of the simulations.
Therefore, we cannot directly compare their outflows properties with the simulated values because of the different ranges of masses or SFRs. 
However, we can see an emerging picture of outflows at $z\simeq2$ if we assume that all these results are roughly following the same trend, within a factor or 2.
SF galaxies at $z\simeq2$, with a SFR higher by a factor 2-4, generate outflows with maximum velocities 2-3 times higher.
The  broad-to-narrow flux ratio is 1.3-1.7 times higher.
More observations of outflows from low-mass ($\Ms \simeq 10^{10} \ \msun$) galaxies along the main-sequence ridge at $z\simeq2$ (${\rm SFR}\simeq30 \sy$) will be needed in order to constrain this trends.

\section{Velocity profiles}
\label{sec:profiles}

In the previous section we identified ionized gas with large velocities perpendicular to the galaxy plane in the tails of the velocity distribution functions (\Fig{Vz_dust}). 
In this section, we are going to study its spatial extent above the disc.
The goal is to identify the nature of these gas motions. 
Do we see inflowing material towards the galaxy plane or outflowing material moving away from the galaxy into the CGM?
Are these gas flows confined within a thin or thick disc? In order to answer these questions
we compute the mean height above (or below) the galaxy plane (Z) in the same velocity bins used in \Fig{Vz_dust}.

In two thirds of the sample we see highly  blueshifted (negative) velocities, coming from material in the near-side of the galaxy (negative Z), upto 8 kpc above the galaxy plane (\Fig{Z_dust}).
This material can be described as large-scale warm outflows with absolute vertical velocities between 100-400 km s$^{-1}$.
At these velocities, the broad component of the velocity distribution function dominates the tail of the distribution.
It accounts for $10^8-10^9 \msun$ of ionized gas, a small but significant fraction of the ISM-CGM medium (deGraf et al., in preparation). We defer the study of the mass outflows rates to that paper.

Lower absolute velocities ($V_{\rm z}= \pm 100 \kms$) are usually confined at $\pm2$ kpc above or below the galaxy plane.
Part of this material could correspond to the outflowing gas being accelerated outwards the disc,
or it could be  part of a fountain flow confined within a given height due to gravity and hydrodynamical balance.
Another possibility is that they are turbulent, non-ordered flows within the violently unstable thick disc \citep{DSC, CDB}.
This low velocity range is dominated by the narrow component of the distribution function.

Finally, high and positive velocities ($V_{\rm z} >100 \kms$) show a diverse distribution. 40\% of the simulations show outflow at the far side of the galaxy (positive Z). These outflows are fainter and less extended than the blueshifted outflows, due to dust obscuration. 
These redshifted outflows are better seen through  galaxy-scale holes of low dust column density,
like in V13. On the other hand,
V19 is an exception, because it shows 
large positive velocities in the near-side of the galaxy, up to Z=-8. This may indicate the presence of extended inflows of gas towards the plane.
However, The majority of the other cases show positive velocities confined to the thick disc. 
This could be an indication of galaxy fountains where high positive velocities appear both in the far-side of the galaxy as outflow and in the near-side of the galaxy plane as inflow.

 \begin{figure}
\includegraphics[width=0.49 \textwidth]{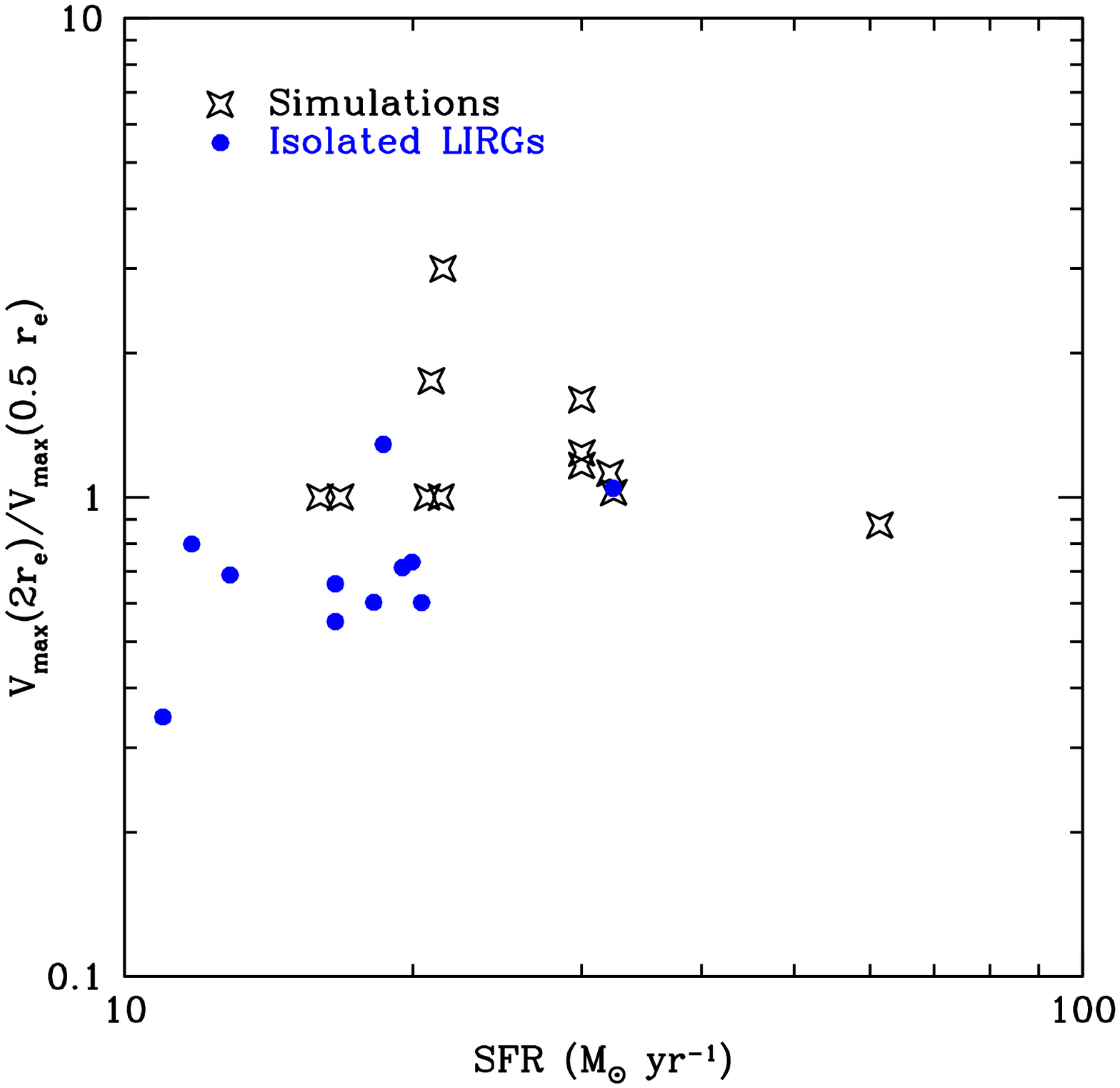}
\includegraphics[width=0.49 \textwidth]{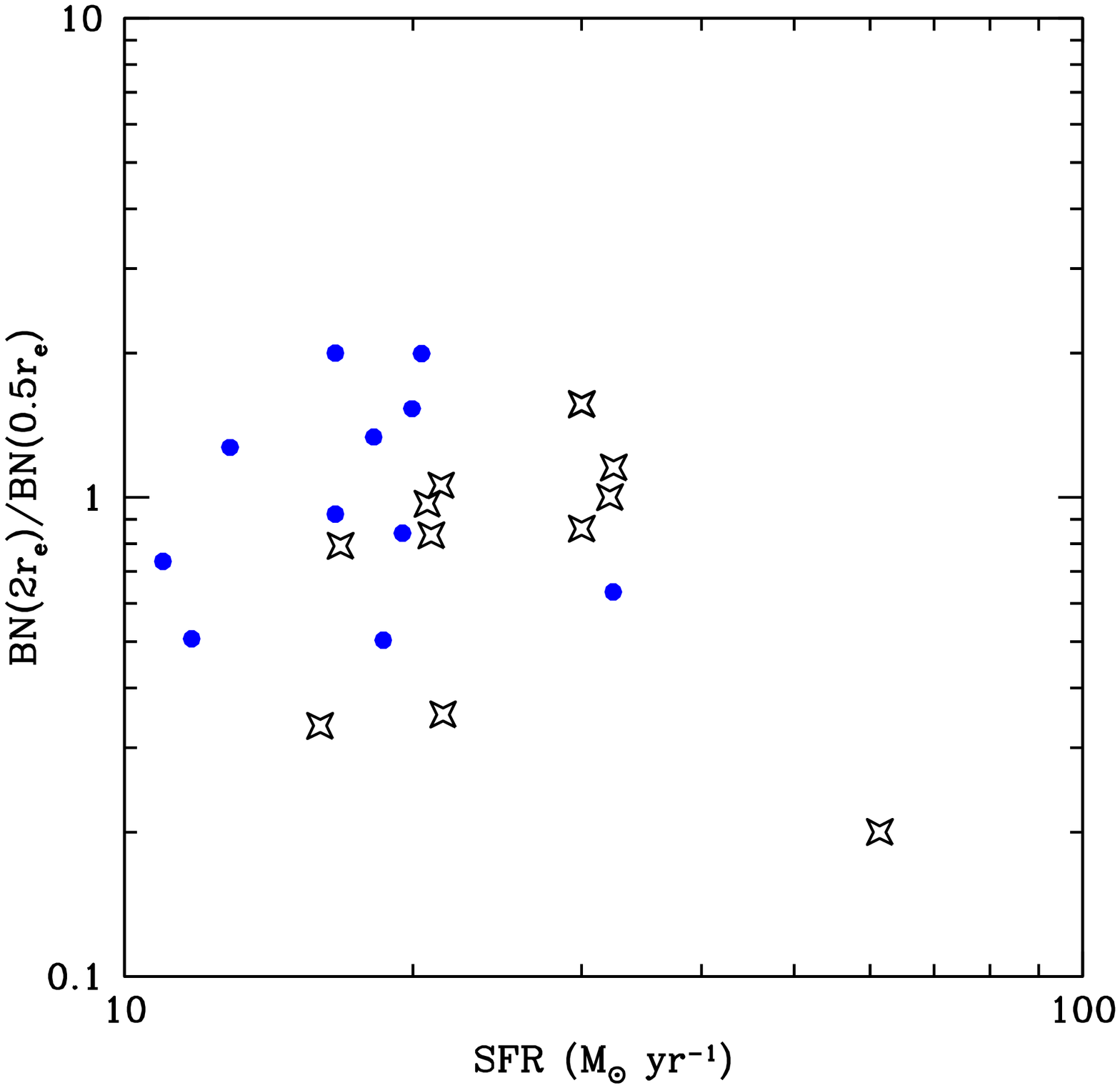}
\caption{Top: Ratio between $V_{\rm max}$ at 2$\re$ and 0.5$\re$ for the simulated galaxies and the sample of local isolated LIRGs of Arribas et al. (2014). Bottom: Ratio of the Broad-to-Narrow flux line ratio.
The simulated and observed values are roughly consistent with each other,
although the observed velocities are slightly higher in the inner parts of galaxies.}
\label{fig:ratioVmax}
\end{figure}

\begin{figure*}
\includegraphics[width=0.99 \textwidth]{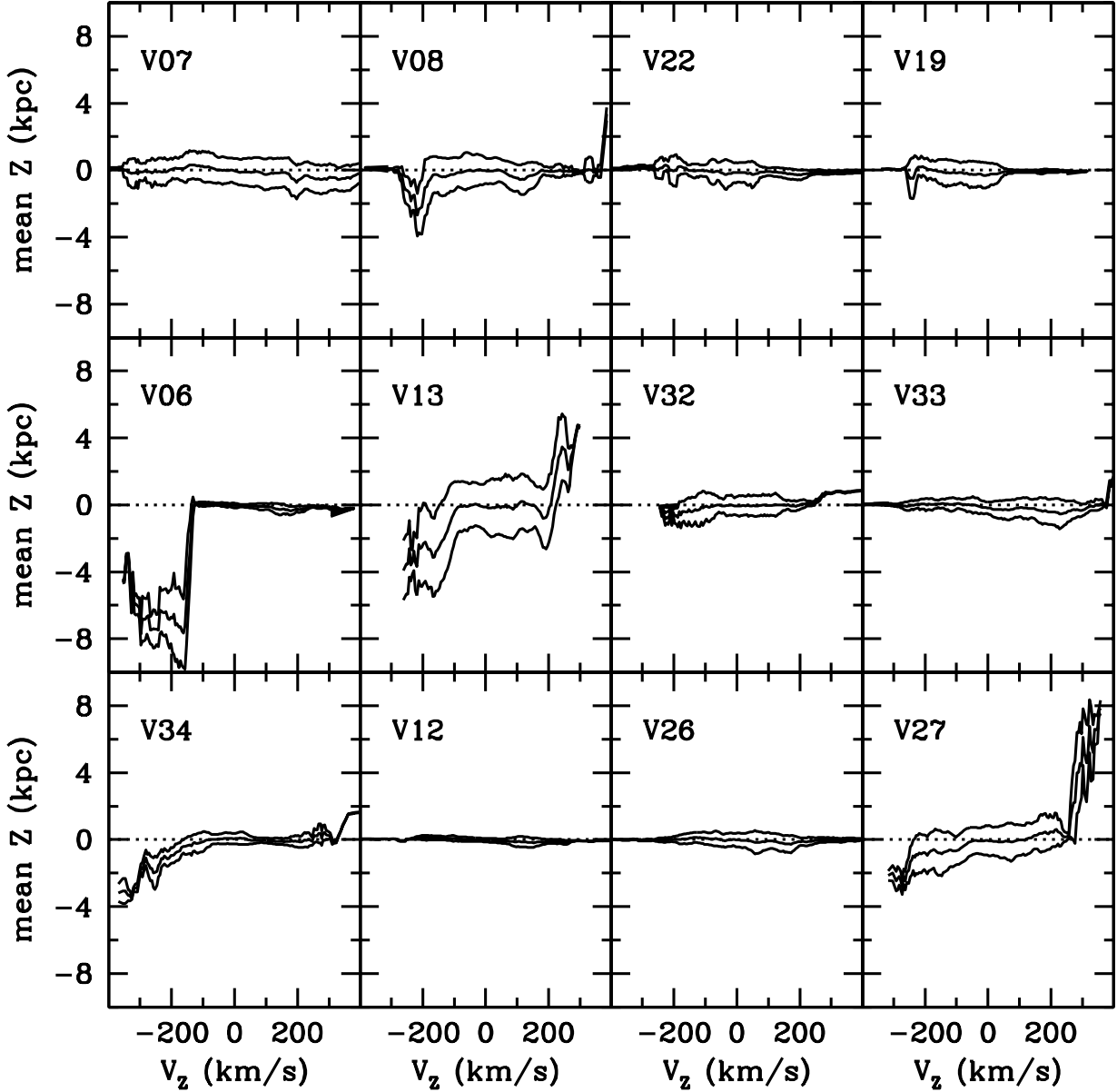}
\caption{Mean and $\pm1\sigma$ dispersion of the height of \Halpha-emitting gas inside $0.5\re$.
The circumnuclear flows are more confined to the galaxy plane than the outer flows, due to the deeper potential perpendicular to the disc.}
\label{fig:Z_nuclear_dust}
\end{figure*}

 \begin{figure*}
\includegraphics[width=0.99 \textwidth]{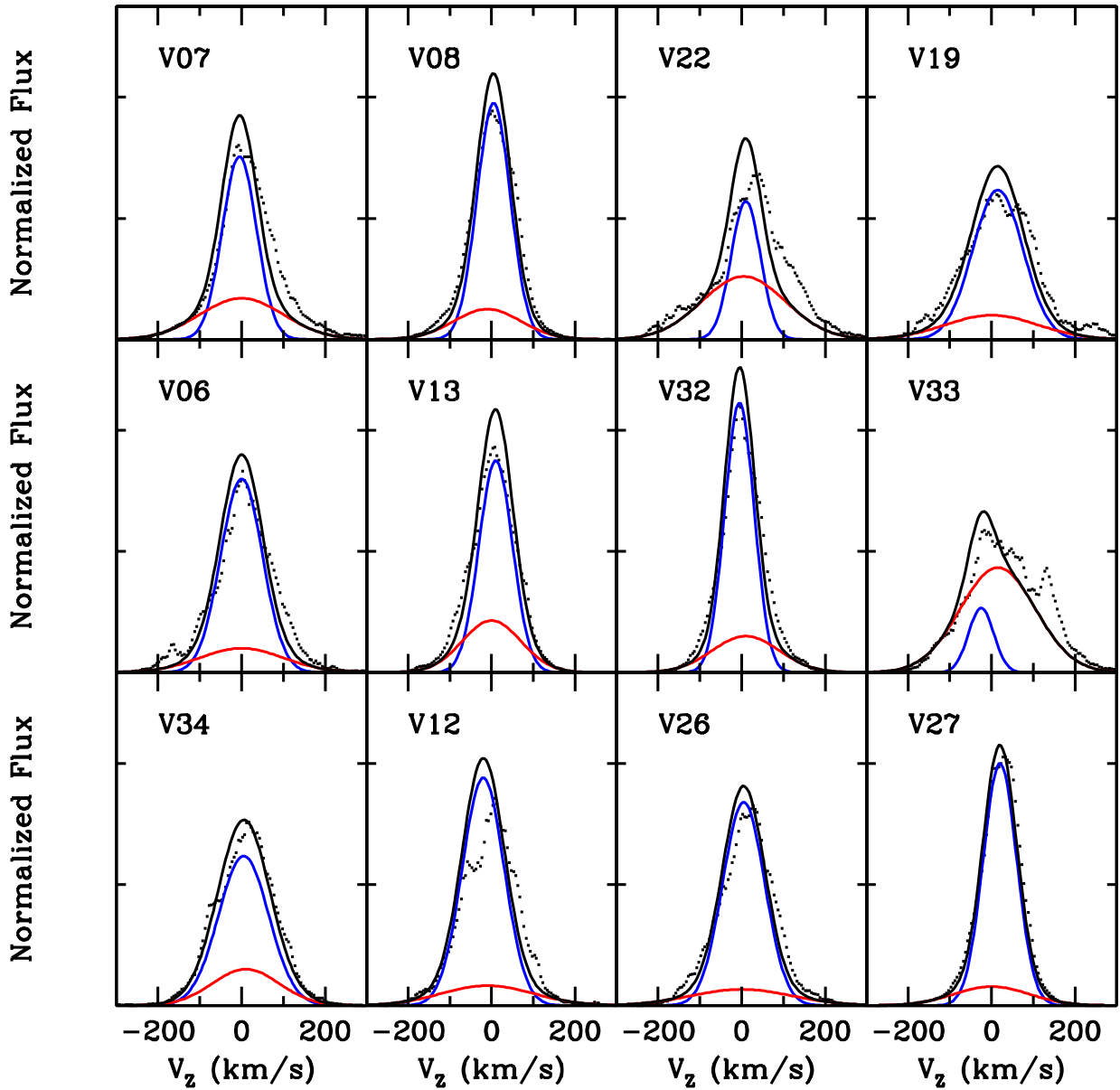}
\caption{Velocity distribution functions with 5 times more dust attenuation (Black points). Curves are the same as in \Fig{Vz_dust}.
The central peak is relatively lower, so that the contribution of the broad component is significantly higher.
However, the high-velocity tails show little differences, so that the maximum velocity of the outflow remains robust against different assumptions about dust attenuation. }
\label{fig:Vz_dust50}
\end{figure*}

\begin{figure*}
\includegraphics[width=0.99 \textwidth]{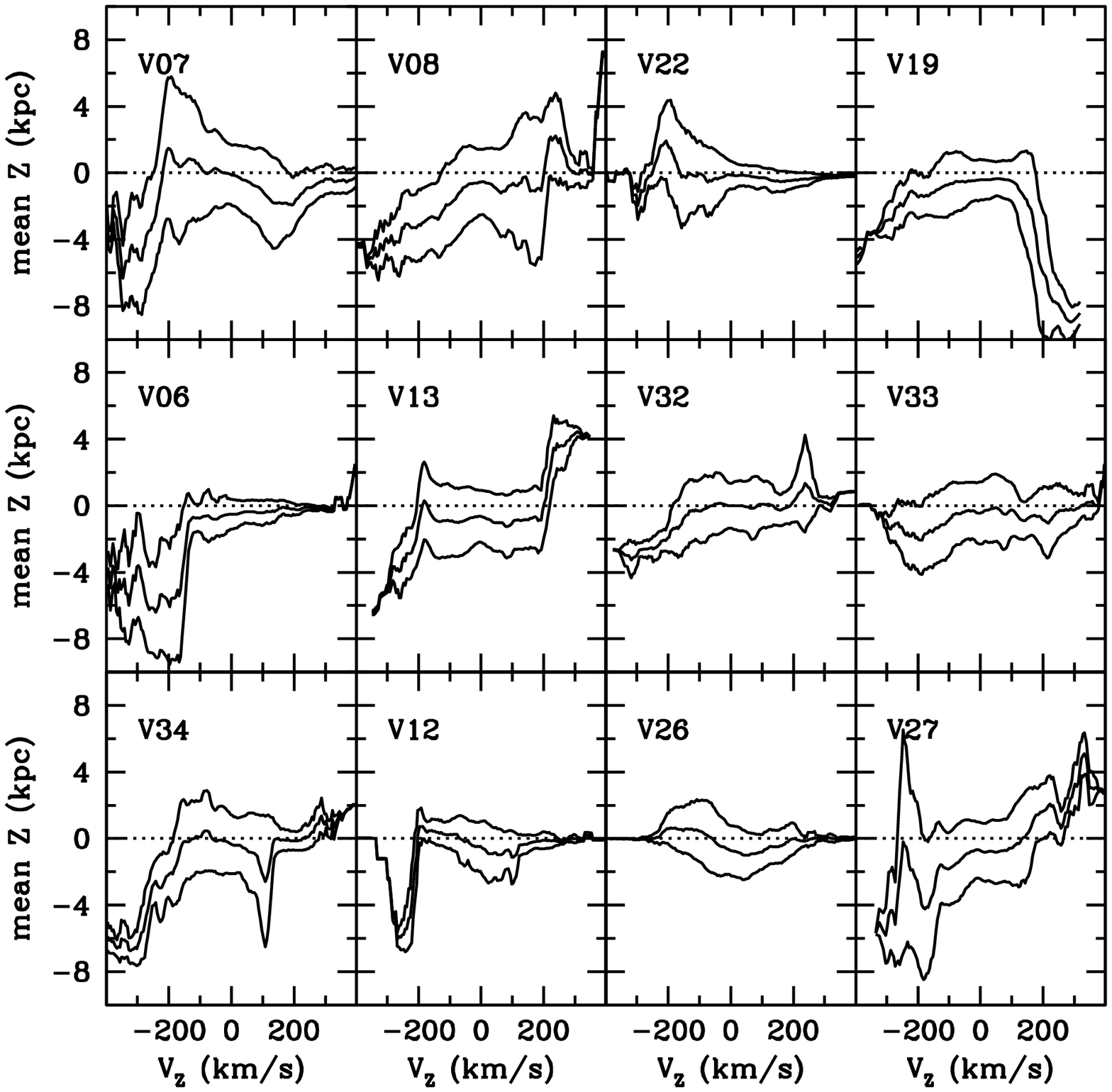}
\caption{Mean and $\pm1\sigma$ heights for different velocities, assuming 5 times more dust attenuation.
The gas in the far-side of the galaxy (Z>0) is more attenuated, so that blueshifted outflows are more visible.}
\label{fig:Z_dust50}
\end{figure*}

If we stack the profiles of all the simulations (\Fig{Z_mean}), we see the combination of a blueshifted outflow at the near-side of the galactic plane plus high velocities of around  $\pm 200 \kms$
confined within $\pm2$ kpc at both sides of the galaxy plane. This could be the signature of galactic fountains.
The mean height ($\bar{Z}$) of the outflow can be fitted with an exponential profile:
\begin{equation} 
\bar{Z} = 1 - \exp ( - ( \bar{V}_z + V_0 ) / w )
\label{eq:fit}
\end{equation}
where $V_0=80 \kms$ is the initial velocity of the outflow at the galaxy plane and $w=190 \kms$ is the characteristic velocity scale.
The above equation is just a fit to the mean outflow profile and its physical meaning is unclear.
However, the profile shows that the outflow accelerates as the material moves away from the galaxy plane.
This profile is inconsistent with homogeneous outflows moving at a constant velocity.
The acceleration is the consequence of having a shock moving in a medium with a declining density profile
\citep[][and references therein]{Chevalier92}.
These shocks could be observed in the outflow spectra as LINER-like features \citep{MonrealIbero06,MonrealIbero10}.

The cartoon models of \Fig{fountain} could help in the interpretation of the negative heights at negative velocities in \Fig{Z_mean}
as indication of an outflow in the near-side of the galaxy.
In a galactic fountain, gas circulates from the galaxy plane ($Z=0$) to a given height above ($Z<0$) or below ($Z>0$) the plane.
In that scenario, material with a given velocity, either positive or negative, is found at both sides of the galaxy plane.
Therefore, the mean height at that velocity is zero and the standard deviation of the height gives the extent of the fountain in the vertical direction. 
Therefore, a set of fountains across the disc would give a mean height equal to zero and the dispersion of $\sigma_Z=2 \kpc$ found in \Fig{Z_mean}. 

An extended outflow in the near-side of the galaxy ($Z<0$) could concentrate gas with blueshifted velocities ($V_Z<0$) in that side of the galaxy. This would give rise to a negative mean height at that velocity, as seen in \Fig{Z_mean}.
These kpc-scale outflows could be observed with the current generation of integral-field spectrographs in local, edge-on LIRGs.

Both scenarios of high vertical velocities are driven by feedback. 
However, random (turbulent) motions in violently unstable thick discs could also give results similar to the fountain scenario at low absolute velocities, where most of the \Halpha \ flux is emitted.
In realistic conditions, the motions  sketched in \Fig{fountain} do not need to be as ordered as ideally represented, but instead they could be more chaotic as a consequence of turbulence and density inhomogeneities in the multiphase medium.

\section{Circumnuclear Outflows}
\label{sec:nuclear}

Previous sections have considered the vertical velocities of the \Halpha-emitting gas integrated over the whole galaxy up to 2$\re$.
Now, we consider only the central regions, up to 0.5$\re$ (0.6-3.7 kpc).
\Fig{Vz_nuclear_dust} shows the distribution function of vertical velocities.
The central peak is slightly lower than before (\Fig{Vz_dust}). 
Therefore, the relative contribution of the narrow component is slightly lower (\tab{2}).
In some cases, the distribution is very broad, like in V07 or V26, making difficult to get some good fits, partially due to the significantly lower number of cells in the nuclear regions.
This broader distributions at the center may be a consequence of the deeper potencial perpendicular to the disc, so the confined fountain flows can move faster if there are in hydrodynamic equilibrium.

\Fig{ratioVmax} shows the ratios between the outflows properties at the two different apertures.
The outflows in the inner regions have similar properties than throughout the galaxy.
The maximum velocity inside 0.5$\re$ is  $V_{\rm max}=100 \pm 40 \kms$, only slightly lower than inside 2$\re$.
On the other hand, the broad-to-narrow flux ratio is slightly higher at the center, F(B)/F(N)=0.4.
\Fig{ratioVmax} also shows the observational results from the sample of \cite{Arribas14}.
Using the same methodology, the outflows properties at two different apertures (2$\re$ and 0.5$\re$) are compared.
Overall, the observed values are roughly (within the scatter) consistent with the simulations.
It seems that the maximum velocities in the inner parts are slightly higher than in the outer parts, so that its median ratio is 
slightly lower than unity,
$V_{\rm max}(2 \re) / V_{\rm max}(0.5 \re) = 0.7$.
More data is needed in order to see stronger trends and/or to reduce the observed scatter.

The inner flows are more confined to the galaxy plane ($\pm 1 \kpc$) than the outer flows (\Fig{Z_nuclear_dust}).
However, 40\% of the sample show kpc-scale outflows. They
spread out to 4 kpc above the galaxy, a factor 2 shorter than in the galaxy-wide outflows discussed in the previous section.
For example, V13 has regions with low \Halpha \ emissivity within $0.5\re$ that could be part of a large kpc-size outflow above the plane.
In other cases (V07), there are no significant outflows in the central region, while a very extended outflow is seen at larger radii,
concentrated in a ring of clumps.

Circumnuclear outflows show similar velocities than the galaxy-wide outflows but their vertical extent is significantly smaller. The deeper potential and higher escape velocity at the center may explain these differences.
V06 is an exception. Its nuclear outflow is equally extended in the vertical direction within 0.7 kpc than within 2.4 kpc.
This galaxy also shows the highest SFR surface density inside $0.5\re$: $\Sigma_{\rm SFR}=4.3 \sy \kpc^{-2}$.
The large release of energy due to this concentration of star-formation could compensate the deeper potential and drive a large-scale outflow
in the inner regions of this galaxy.

\section{The effect of increasing dust attenuation}
\label{sec:dust}

 The distribution of dust  and its effect in the modification of the \Halpha \ line is an important uncertainty in this study.
 Therefore, we perform the experiment of increasing the dust optical depth by a factor of 5, mimicking the effect of a higher dust opacity.
 or a higher amount of dust. This increases the \Halpha \ attenuation in $1.7 \pm 0.3$ magnitudes, depending on the galaxy.
 
 \Fig{Vz_dust50} shows the velocity distributions similar to \Fig{Vz_dust}.
 The central peak is relatively lower
than in the case with less attenuation. This means that the contribution of the broad component is higher than before.
 The median broad-to-narrow ratio increases to F(B)/F(N)=0.5 $\pm 0.2$, excluding the values higher than unity  (\tab{2}).
 The difference is 70\% of the fiducial value, still within the 2$\sigma$ dispersion of the sample.
 Comparing  with the results of \se{Vz}, these higher predicted F(B)/F(N) values agree better with those observed  for the LIRGs sample, for which this higher attenuation values have been reported \citep{Veilleux99, PiquerasLopez13}.  
 On the other hand, the broad component is overall unchanged, $V_{\rm max}=110 \pm 20 \kms$.
 Therefore, we conclude that the different assumptions about dust attenuation change the distributions of vertical velocities, but the changes are constrained to the dense and turbulent ISM (narrow component).

The mean height above or below the galaxy plane are qualitatively unchanged by using different dust assumptions (\Fig{Z_dust50}).
However, the material at the far-side of the galaxy (Z>0) are significantly more attenuated, so the mean Z values are slightly shifted towards more negative values (towards the observer). That shift is small, less than 1 kpc in most cases.
Exceptional cases are V27 and V08, where material far away from the plane (Z=2-4 kpc) are severely attenuated.
This makes the near-side outflows (Z<0) more prominent because they are less attenuated.
We conclude that more dust attenuation makes the near-side, blueshifted outflows more visible, by the higher attenuation of 
material with similar velocity along the same line-of-sight but located in the other side of the galaxy.
 
\section{Conclusion}
\label{sec:conclusion}

We addressed the velocities of warm galactic outflows from synthetic \Halpha \ observations of star-forming, non-merging galaxies from the \textsc{Vela} sample of zoom-in AMR cosmological simulations of galaxy formation.
The main conclusions can be summarized as follows

\begin{itemize}
\item
\Halpha \ emitting gas is coming from star-forming regions (clumps)
within a violently unstable disc but also from diffuse,
$10^4$K-degrees, ionized gas around the galaxy.
\item
The computed \Halpha \ luminosity agrees with the expected \Halpha \
luminosity based on the galaxy SFR \citep{Kennicutt98}, after assuming
a spatially constant clumping factor of about 24.
\item
The distribution function of velocities perpendicular to the galaxy plane can be described as the sum of two gaussians: a narrow component (2/3 of the total flux) plus a broad component, both centered around the galaxy systemic velocity. 
\item
The narrow components have a median velocity width of $\sigma_{\rm N}=40 \pm 10 \kms$ and they could be identified as the internal (turbulent) motions within the violently unstable discs.
\item
The broad components have a maximum velocity of $ V_{\rm max} = 110 \pm 20 \kms$ and they could be identified as part of a galactic outflow or a galactic fountain.
\item
The properties of the synthetic \Halpha \ emission lines from simulated galaxies at $z\simeq2$ are in general good agreement with those observed in local isolated galaxies of similar masses and SFRs \citep{Arribas14}.  
\item
Two thirds of the sample show spatially extended, blueshifted outflows.
This material is coming from the near-side of the galaxy upto Z=8 kpc above the galaxy plane.
\item
Outflowing gas accelerates as it leaves the galaxy plane. 
This is inconsistent with the scenario of outflows moving with a constant, space-independent velocity.
\item
Outflows are spread over the whole galaxy, and they show similar high velocities.
However, the vertical extent of the circumnuclear outflows
is smaller than the outflows in the outer parts of galaxies.
\item
Assuming five times higher dust attenuation does not change the properties of the outflows.
\end{itemize}

In this paper, we consider the velocities of outflows perpendicular to the galaxy plane.
We found that a small although significant fraction of the ionized, warm gas escapes the disc and joins the CGM in some cases or
it remains confined within a thick disc, as part of galactic fountains or in a turbulent medium.
However, this vertical geometry not longer applies if we consider larger radii away from the disc.
Outside galactic discs, radial flows could propagate outwards, even along the galaxy plane.
This and other issues related with outflows in the CGM will be discussed in a companion paper (deGraf. et al., in preparation).

\section*{Acknowledgments} 

The simulations were performed 
at the National Energy Research Scientific Computing Center (NERSC) at  
Lawrence Berkeley National Laboratory, and 
at NASA Advanced Supercomputing (NAS) at NASA Ames Research Center.
Development and most of the analysis have been performed in the astro cluster at HU.
This work has been partly funded by the Spanish Ministry of 
 Economy and Competitiveness, project  AYA2012-32295
and by the ERC Advanced Grant, STARLIGHT: Formation of the First Stars (project number 339177).

\bibliographystyle{mn2e}
\bibliography{winds3}

\bsp

\label{lastpage}

\end{document}